\documentclass[%
superscriptaddress,
 amsmath,amssymb,
prx,    
floatfix,
twocolumn
]{revtex4-2}

\usepackage{physics}
\usepackage{graphicx}
\usepackage{dcolumn}
\usepackage{bm}
\usepackage{hyperref}
\hypersetup{colorlinks=true,citecolor={blue},linkcolor={blue},urlcolor={blue}}

\newcommand{\bk}{\mathbf{k}}
\newcommand{\bp}{\mathbf{p}}
\newcommand{\bq}{\mathbf{q}}

\newcommand{\cg}{\mathcal{G}}
\newcommand{\cf}{\mathcal{F}}

\usepackage{tikz}
\usetikzlibrary{decorations.pathmorphing}
\tikzset{snake it/.style={decorate, decoration=snake}}
\usetikzlibrary{decorations.markings}
\tikzset{middlearrow/.style={ decoration={markings, mark= at position 0.575 with {\arrow{#1}}, }, postaction={decorate} } }

\begin{document}

\title{Strong-coupling theory of condensate-mediated superconductivity in 2D materials}


\author{Meng Sun}
\affiliation{Center for Theoretical Physics of Complex Systems, Institute for Basic Science (IBS), Daejeon 34126, Korea}

\author{A.~V.~Parafilo}
\affiliation{Center for Theoretical Physics of Complex Systems, Institute for Basic Science (IBS), Daejeon 34126, Korea}

\author{V.~M.~Kovalev}
\affiliation{A.V.~Rzhanov Institute of Semiconductor Physics, Siberian Branch of Russian Academy of Sciences, Novosibirsk 630090, Russia}
\affiliation{Novosibirsk State Technical University, Novosibirsk 630073, Russia}

\author{I.~G.~Savenko}
\affiliation{Center for Theoretical Physics of Complex Systems, Institute for Basic Science (IBS), Daejeon 34126, Korea}
\affiliation{Basic Science Program, Korea University of Science and Technology (UST), Daejeon 34113, Korea}

\date{\today}

\begin{abstract}
We develop a strong-coupling theory of Bose-Einstein condensate-mediated superconductivity in a hybrid system, which consists of a two-dimensional electron gas with either (i) parabolic spectrum or (ii) relativistic Dirac spectrum in the vicinity of a two-dimensional solid-state condensate of indirect excitons. 
The Eliashberg equations are derived and the expressions for the electron pairing self-energy due to the exchange interaction between electrons mediated by a single Bogoliubov excitation (a bogolon) and the bogolon pairs are found. 
Furthermore, we find the superconducting order parameter and estimate the critical temperature of the superconducting transition. 
The critical temperature reveals its linear dependence on the dimensionless coupling constant.
It is shown, that the bogolon-pair-mediated interaction is the dominant mechanism of electron pairing in hybrid systems in both the weak and strong coupling regimes.
We calculate the effective bogolon-electron interaction constant for both parabolic and linear electron dispersions and examine the dependence of the critical temperature of electron gas superconducting transition on exciton condensate density.
\end{abstract}

\maketitle


\section{Introduction}
The first microscopic description of the superconducting (SC) state belongs to the celebrated Bardeen-Copper-Schrieffer (BCS) theory~\cite{RefBCS,PhysRev.108.1175,RevModPhys.62.1027,AllenBook}.
Using this theory, one can explain the emergence of the SC gap and estimate the critical temperature of SC transition $T_c$.
Later, Green's functions formalism-based Migdal-Eliashberg (later Eliashberg) theory was developed~\cite{migdal1958interaction,RefEliashberg1,RefEliashberg2}.
It provides a more rigorous and accurate basis for the estimation of $T_c$ and allows to account for the Coulomb repulsion between electrons.
Generally, by employing the McMillan's approach~\cite{PhysRev.167.331}, the formation of Cooper pairs by electron-phonon interaction can be solely determined by the effective electron-phonon coupling strength $\lambda$. 
As compared with the BCS theory, which only works in the weak coupling regime (of $\lambda\ll1$), the Eliashberg theory is more general.
In particular, it avoids the Debye frequency cutoff, thus allowing to extend the estimations to the strong coupling regime (of $\lambda$ close to or larger than unity).

The price to pay is that the analytical and numerical calculations by the Eliashberg theory are usually more complicated than the ones by the BCS. It requires the solution of multiple coupled equations, which is often tricky, and some assumptions are in order.
Since 1960s, various numerical approaches implying different special assumptions and simplifications have been developed in order to find the approximate solutions of the Eliashberg equations.
Very recently, there have been suggested powerful density functional-based techniques (called the density functional theory for superconductors)~\cite{PhysRevB.72.024545, PhysRevB.72.024546,PhysRevLett.125.057001}. 
In general, the phonon-mediated superconductivity is well studied, though there still exist many open questions, in particular, regarding the SC transition in novel two-dimensional (2D) materials and their stacks due to phonon-mediated pairing.

However, electron-phonon interaction is not the only possible route to the formation of Cooper pairs.
Among others, there have been reported several proposals for exciton~\cite{PhysRev.134.A1416,ginzburg,PhysRevB.7.1020}, exciton-polariton~\cite{Laussy:2010aa,kavokin2016exciton,Skopelitis:2018aa}, and cavity-photon-mediated~\cite{PhysRevLett.122.133602} coupling mechanisms between electrons.
In our recent works~\cite{Sun21:th,sun2020boseeinstein}, we have developed a BCS-like theory for the indirect exciton condensate-mediated superconductivity in hybrid systems consisting of layers of a (Bose-condensed) exciton gas and a two-dimensional electron gas or graphene.
Employing the Bogoliubov theory, we considered quasiparticle excitations above the exciton condensate called bogolons. 
They interact with the electron gas via Coulomb forces, which distinguishes this problem from the phonon-assisted electron-electron interaction in conventional superconductors. 

However, all the proposals listed above are based on the BCS-like approach. 
Thus, the weak coupling between electrons and phonons (or other particles like bogolons) is assumed, which means that $\lambda$ should, strictly speaking, be small.
In this paper, we build a strong-coupling theory for bogolon-mediated superconducting in hybrid Bose-Fermi systems.
By employing the Green's functions technique, we calculate the electron-bogolon coupling strength $\lambda$ and give an estimation for the critical temperature of SC transition.
The bosonic subsystem which we consider here is indirect exciton gas. 
It can be substituted by other quasi-condensates like direct exciton gas or exciton-polariton condensates, which have been predicted~\cite{imamog1996nonequilibrium,lozovik,Fogler2014,PhysRevB.92.165121, PhysRevB.93.245410,PhysRevB.96.174504} and studied experimentally~\cite{Lerario:2017aa,Room1, Room2, WangNature2019}.
The condensation in these systems has been reported at relatively high temperatures, even approaching the room temperature.

The paper is organized as follows. 
In Sec.~\ref{sec:interaction}, we introduce the  Hamiltonian of bogolon-electron interaction.
In Sec.~\ref{sec:etheory}, we build the bogolon-mediated Eliashberg theory and derive the Eliashberg spectral function.
Furthermore, in Sec.~\ref{sec:res}, we calculate the electron-bogolon coupling constant $\lambda$ and estimate the critical temperature by considering electrons with parabolic and linear dispersions.
Finally, in Sec.~\ref{sec:res} we summarize the results.


\section{Single-bogolon and bogolon-pair-mediated interaction} \label{sec:interaction}
\begin{figure}[!t]
\includegraphics[width=0.45\textwidth]{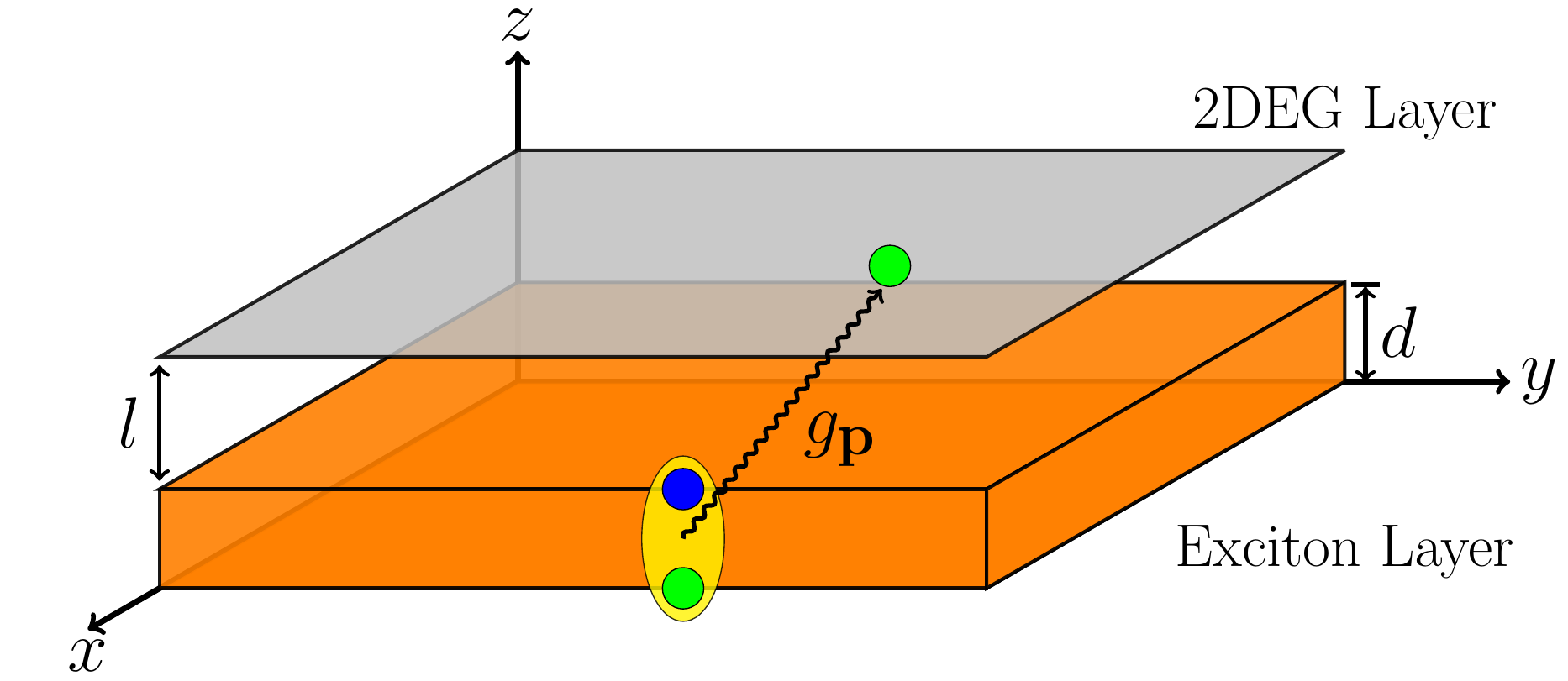}
\caption{System schematic.
Two-dimensional electron gas (2DEG) in the vicinity of a 2D Bose-Einstein condensate (BEC).
We consider the BEC of indirect excitons (yellow), which reside in a double quantum well: n-doped and p-doped layers of MoS$_2$ and WSe$_2$  separated by an hBN.
Electrons and the condensate particles are coupled via the Coulomb forces $g_\mathbf{p}$, which allows electrons with opposite spins form Cooper pairs.
}
\label{Fig1}
\end{figure}

Let us consider a hybrid system, in which a layer of electron gas (2DEG) and a layer of a Bose-Einstein condensate are in the vicinity of each other in $z$-direction (Fig.~\ref{Fig1}).
For the BEC part, we consider indirect excitons, where the formation of a BEC has been recently reported~\cite{Butov2017, WangNature2019}.
Indirect excitons consist of electrons and holes residing in n- and p-doped layers which are separated in $z$ direction.
These layers can be made of, e.g., MoS$_2$ or WSe$_2$ materials separated by several layers of hexagonal boron nitride (hBN)~\cite{WangNature2019,backes2020production}.
The 2D electron gas (described by the field operator $\Psi_{\sigma}(\mathbf{r})$ with the position vector $\mathbf{r}$ and the spin $\sigma=\{\uparrow,\downarrow\}$) and the exciton gas (described by the field operator $\Phi(\mathbf{r})$) layers are also spatially separated by a hBN layer.
The two types of particles, electrons and indirect excitons, are coupled through the Coulomb interaction (in standard form)~\cite{Boev:2016aa, Matuszewski:2012aa}, 
\begin{equation}\label{eq.1}
{\cal H}=\sum_{\sigma}\int d\mathbf{r}\int d\mathbf{R}\Psi^\dag_{\sigma}(\mathbf{r})\Psi_{\sigma}(\mathbf{r})g\left(\mathbf{r}-\mathbf{R}\right)\Phi^\dag(\mathbf{R})\Phi(\mathbf{R}).
\end{equation}
Here, $\bold{R}$ is the in-plane position vector of the exciton center-of-mass position. 
We consider the case when the most of  excitons are being in the ground state (in BEC). 
Then, in the weakly-interacting regime, we can write the exciton field operator as $\Phi(\mathbf{R})=\sqrt{n_c}+ \varphi(\mathbf{R})$, where $n_c$ is the condensate density and $\varphi(\mathbf{R})$ is the field operator for non--condensed excitons. Applying the Fourier and the Bogoliubov transformations, from~\eqref{eq.1}, we can find the Hamiltonian for the one-bogolon and bogolon-pair-mediated interactions~\cite{Sun2018,Villegas2019,sun2020boseeinstein,Sun21:th} (putting $\hbar = k_B =1$ below for simplicity and restoring these constants later)
%
\begin{eqnarray}
\label{Eq1bexpr}
&&{\cal H}_1= \frac{\sqrt{n_c}}{L} \sum_{\bf k p \sigma} g_{\bf p} \left[(v_{\bf p}+u_{\bf -p})b^{\dag}_{-\bf p} \right. \\ \nonumber
&&\left.~~~~~~~~~~~~~~~~~~~~+(v_{\bf -p}+u_{\bf p})b_{\bf p}\right]c^{\dag}_{\bf k+p,\sigma}c_{\bf k,\sigma},\\
\label{Eq2bexpr}
&&{\cal H}_2= \frac{1}{L^2} \sum_{\bf k p q \sigma} g_{\bf p} \left[ u_{\bf q-p}u_{\bf q}b^{\dag}_{\bf q-p}b_{\bf q}+u_{\bf q-p}v_{\bf q}b^{\dag}_{\bf q-p}b^{\dag}_{\bf -q}\right. \nonumber \\ 
&&\left.+v_{\bf q-p}u_{\bf q}b_{\bf -q+p}b_{\bf q}
+v_{\bf q-p}v_{\bf q}b_{\bf -q+p}b^{\dag}_{\bf -q}
\right]
c^{\dag}_{\bf k+p,\sigma}c_{\bf k,\sigma},
\end{eqnarray}
where $g_\mathbf{p}$ is the Fourier image of electron-exciton interaction;
$c_{\mathbf{p},\sigma}$ and $b_\mathbf{p}$ are the annihilation operators for the electrons and bogolons, respectively.
The Bogoliubov coefficients are defined as~\cite{Giorgini:1998aa}
\begin{eqnarray}
&&u^2_{\mathbf{p}}=1+v^2_{\mathbf{p}}=\frac{1}{2}\left(1+\left[1+\left(\frac{Ms^2}{\omega_{\mathbf{p}}}\right)^2\right]^{1/2}\right),\label{eq.4-1}
\\
&&~~~~~u_{\mathbf{p}}v_{\mathbf{p}}=-\frac{Ms^2}{2\omega_{\mathbf{p}}},\label{eq.4-2}
\end{eqnarray}
where $M$ is the exciton effective mass, $s=\sqrt{\kappa n_c/M}$ is the sound velocity, $\kappa=e_0^2d/\epsilon_0\epsilon$ is the exciton-exciton interaction strength in the reciprocal space, $e_0$ is electron charge, $\epsilon$ is the dielectric constant, $\epsilon_0$ is the dielectric permittivity, and
$\omega_p= sp(1+p^2\xi_h^2)^{1/2}$ is the spectrum of bogolons with the definition of healing length $\xi_h=1/2Ms$.

To get an analytic form of the electron-exciton interaction, we disregard the peculiarities of the exciton internal motion (relative motion of the electron and hole in the exciton). 
In monolayers of transition metal dichalcogenides, the exciton binding energy is usually very large: it might even exceed the room temperature. 
Thus, the assumption that an individual exciton is in its ground state with respect to its relative electron-hole motion is  reasonable, and only the exciton center-of-mass motion plays an important role.
Then, the electron-exciton interaction in direct space reads
\begin{eqnarray}
\label{EqgDirSp}
g(\mathbf{r} - \mathbf{R} ) &=& \frac{e_0^2}{4\pi \epsilon_0 \epsilon} \left( \frac{1}{r_{ee}} - \frac{1}{r_{eh}} \right),
\end{eqnarray}
with $r_{ee}=\sqrt{l^2 +(\mathbf{r}-\mathbf{R})^2}$  and $r_{eh}= \sqrt{(l+d)^2 + (\mathbf{r}-\mathbf{R})^2}$;
here, $d$ is an effective size of the boson, which is equal to the distance between the n- and p-doped layers in the case of indirect exciton condensate, and $l$ is the separation between the 2DEG and the BEC as shown in Fig.~\ref{Fig1}.
Performing the Fourier transformation, we find
\begin{equation}
 g_p=\frac{e^2_0 (1-e^{-pd})e^{-pl}}{2\epsilon_0\epsilon p}. \label{gp}
\end{equation}

Using the conventional perturbation theory, one may argue that $\mathcal{H}_1\gg \mathcal{H}_2$ since the density of non-condensed particles is a small quantity as compared with the condensate density, i.e. $\sqrt{n_c}b^\dagger_{\mathbf{p}}\gg b^\dagger_\mathbf{p}b_\mathbf{p}$.
However, with the Bogoliubov coefficients, the summation such as $v_\mathbf{p}+u_{-\mathbf{p}}$ in \eqref{Eq1bexpr} can drastically change the situation.
In the long wavelength limit, $\xi_h p \ll 1$, where we are interested in this work, we have $\omega_p \approx sp$ and $u_\mathbf{p} \approx -v_\mathbf{p} \approx \sqrt{Ms/p}$.
Then, in this limit, one has $v_\mathbf{p}+ u_{\mathbf{p}} \to 0$ whereas $u_\mathbf{p}v_\mathbf{p}\to \infty$.
This suggest the contribution of $\mathcal{H}_2$ may surpass the one in $\mathcal{H}_1$ as will be shown in Sec.~\ref{sec:etheory}.


%
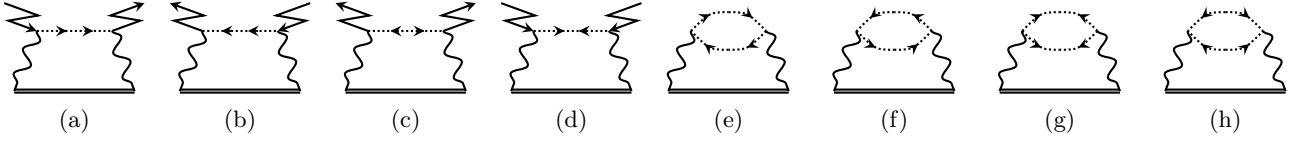
\begin{figure*}[ht!]
\begin{center}
\tikzset{>=stealth}
\tikzset{every picture/.style={line width=0.8pt}}
\def\x{0.4}
\begin{tabular}{ m{0.115\textwidth} m{0.115\textwidth} m{0.115\textwidth} m{0.115\textwidth} m{0.115\textwidth} m{0.115\textwidth} m{0.115\textwidth} m{0.115\textwidth} }
\begin{tikzpicture}
\draw [->] (0.39,0) -- (0.4,0);
\draw [->] (0.74,0) -- (0.75,0);
\draw [densely dotted] (0,0) -- (1,0);
\draw[] (-0.3,-0.78) -- (1.3,-0.78);
\draw[] (-0.3,-0.82) -- (1.3,-0.82);
\draw [snake it] (0,0) -- (-0.3,-0.8);
\draw [snake it] (1,0) -- (1.3,-0.8);
\draw [<-] (0,0) --++ ({-\x*cos(20)},{\x*sin(20)}) -- ++ ({\x*cos(10)},{\x*sin(10)}) -- ++ ({-1.2*\x*cos(20)},{1.2*\x*sin(20)});
\draw [->] (1,0) -- ++ ({\x*cos(20)},{\x*sin(20)}) -- ++ ({-\x*cos(10)},{\x*sin(10)}) -- ++ ({1.2*\x*cos(20)},{1.2*\x*sin(20)});
\node at (0.5,-1.2) {(a)};
\end{tikzpicture}
&
\begin{tikzpicture}
\draw [-<] (0.39,0) -- (0.4,0);
\draw [-<] (0.74,0) -- (0.75,0);
\draw [densely dotted] (0,0) -- (1,0);
\draw[] (-0.3,-0.78) -- (1.3,-0.78);
\draw[] (-0.3,-0.82) -- (1.3,-0.82);
\draw [snake it] (0,0) -- (-0.3,-0.8);
\draw [snake it] (1,0) -- (1.3,-0.8);
\draw [->] (0,0) --++ ({-\x*cos(20)},{\x*sin(20)}) -- ++ ({\x*cos(10)},{\x*sin(10)}) -- ++ ({-1.2*\x*cos(20)},{1.2*\x*sin(20)});
\draw [<-] (1,0) -- ++ ({\x*cos(20)},{\x*sin(20)}) -- ++ ({-\x*cos(10)},{\x*sin(10)}) -- ++ ({1.2*\x*cos(20)},{1.2*\x*sin(20)});
\node at (0.5,-1.2) {(b)};
\end{tikzpicture}
&
\begin{tikzpicture}
\draw [<-] (0.3,0) -- (0.31,0);
\draw [->] (0.74,0) -- (0.75,0);
\draw [densely dotted] (0,0) -- (1,0);
\draw[] (-0.3,-0.78) -- (1.3,-0.78);
\draw[] (-0.3,-0.82) -- (1.3,-0.82);
\draw [snake it] (0,0) -- (-0.3,-0.8);
\draw [snake it] (1,0) -- (1.3,-0.8);
\draw [->] (0,0) --++ ({-\x*cos(20)},{\x*sin(20)}) -- ++ ({\x*cos(10)},{\x*sin(10)}) -- ++ ({-1.2*\x*cos(20)},{1.2*\x*sin(20)});
\draw [->] (1,0) -- ++ ({\x*cos(20)},{\x*sin(20)}) -- ++ ({-\x*cos(10)},{\x*sin(10)}) -- ++ ({1.2*\x*cos(20)},{1.2*\x*sin(20)});
\node at (0.5,-1.2) {(c)};
\end{tikzpicture}
&
\begin{tikzpicture}
\draw [->] (0.4,0) -- (0.41,0);
\draw [<-] (0.6,0) -- (0.61,0);
\draw [densely dotted](0,0) -- (1,0);
\draw[] (-0.3,-0.78) -- (1.3,-0.78);
\draw[] (-0.3,-0.82) -- (1.3,-0.82);
\draw [snake it] (0,0) -- (-0.3,-0.8);
\draw [snake it] (1,0) -- (1.3,-0.8);
\draw [<-] (0,0) --++ ({-\x*cos(20)},{\x*sin(20)}) -- ++ ({\x*cos(10)},{\x*sin(10)}) -- ++ ({-1.2*\x*cos(20)},{1.2*\x*sin(20)});
\draw [<-] (1,0) -- ++ ({\x*cos(20)},{\x*sin(20)}) -- ++ ({-\x*cos(10)},{\x*sin(10)}) -- ++ ({1.2*\x*cos(20)},{1.2*\x*sin(20)});
\node at (0.5,-1.2) {(d)};
\end{tikzpicture}
&
\begin{tikzpicture}
\draw [densely dotted, middlearrow=stealth] (0,0) .. controls (0.25,0.25) .. (0.5,0.25);
\draw [densely dotted, middlearrow=stealth] (0.5,0.25) .. controls (0.75,0.25) .. (1,0);
\draw [densely dotted, middlearrow=stealth] (0.5,-0.25) .. controls (0.25,-0.25) .. (0,0);
\draw [densely dotted, middlearrow=stealth] (1,0) .. controls (0.75,-0.25) .. (0.5,-0.25);
\draw[] (-0.3,-0.78) -- (1.3,-0.78);
\draw[] (-0.3,-0.82) -- (1.3,-0.82);
\draw [snake it] (0,0) -- (-0.3,-0.8);
\draw [snake it] (1,0) -- (1.3,-0.8);
\node at (0.5,-1.2) {(e)};
\end{tikzpicture}
&
\begin{tikzpicture}
\draw [densely dotted, middlearrow=stealth] (0.5,0.25) .. controls (0.25,0.25) .. (0.0,0);
\draw [densely dotted, middlearrow=stealth] (1,0) .. controls (0.75,0.25) .. (0.5,0.25);
\draw [densely dotted, middlearrow=stealth] (0,0) .. controls (0.25,-0.25) .. (0.5,-0.25);
\draw [densely dotted, middlearrow=stealth] (0.5,-0.25) .. controls (0.75,-0.25) .. (1,0);
\draw[] (-0.3,-0.78) -- (1.3,-0.78);
\draw[] (-0.3,-0.82) -- (1.3,-0.82);
\draw [snake it] (0,0) -- (-0.3,-0.8);
\draw [snake it] (1,0) -- (1.3,-0.8);
\node at (0.5,-1.2) {(f)};
\end{tikzpicture}
&
\begin{tikzpicture}
\draw [densely dotted, middlearrow=stealth] (0,0) .. controls (0.25,0.25) .. (0.5,0.25);
\draw [densely dotted, middlearrow=stealth] (1.0,0) .. controls (0.75,0.25) .. (0.5,0.25);
\draw [densely dotted, middlearrow=stealth] (0.0,0) .. controls (0.25,-0.25) .. (0.5,-0.25);
\draw [densely dotted, middlearrow=stealth] (1,0) .. controls (0.75,-0.25) .. (0.5,-0.25);
\draw[] (-0.3,-0.78) -- (1.3,-0.78);
\draw[] (-0.3,-0.82) -- (1.3,-0.82);
\draw [snake it] (0,0) -- (-0.3,-0.8);
\draw [snake it] (1,0) -- (1.3,-0.8);
\node at (0.5,-1.2) {(g)};
\end{tikzpicture}
&
\begin{tikzpicture}
\draw [densely dotted, middlearrow=stealth] (0.5,0.25) .. controls (0.25,0.25) .. (0,0);
\draw [densely dotted, middlearrow=stealth] (0.5,0.25) .. controls (0.75,0.25) .. (1.0,0);
\draw [densely dotted, middlearrow=stealth] (0.5,-0.25) .. controls (0.25,-0.25) .. (0.0,0);
\draw [densely dotted, middlearrow=stealth] (0.5,-0.25) .. controls (0.75,-0.25) .. (1,0);
\draw[] (-0.3,-0.78) -- (1.3,-0.78);
\draw[] (-0.3,-0.82) -- (1.3,-0.82);
\draw [snake it] (0,0) -- (-0.3,-0.8);
\draw [snake it] (1,0) -- (1.3,-0.8);
\node at (0.5,-1.2) {(h)};
\end{tikzpicture}
\end{tabular}
\caption{Electron self-energy diagrams. 
The double solid lines stand for the Green's functions of an electron $\hat{G}$ in \eqref{EqGreeenExpl}. 
The zigzag lines denote the condensate particles. 
(Thus, each zigzag line gives the factor $\sqrt{n_c}$). 
The dotted lines represent bogolons (the particles excited from the condensate). 
The wiggly lines stand for the electron-exciton interaction $g_\mathbf{p}$. 
Panels (a-d) correspond to 1b process in~\eqref{self1b}, and panels (e-h) correspond to 2b process in~\eqref{self2b}. Physically, the diagrams in (a-d) describe the excitation of condesate particle to the non-condesed state by a moving electron, whereas (e-h) describe the condensate polarization due to the moving electron.}
\label{fig:diagram}
\end{center}
\end{figure*}
\section{The Eliashberg equations}\label{sec:etheory}
We will work in the framework of the Nambu-Gor'kov formalism~\cite{RefNambu,gor1958energy}.
First, let us introduce the two-component field operator,
\begin{equation} \label{psi_def}
{\bm \Psi}_\bk =
\begin{pmatrix}
c_{\bk \uparrow}\\ 
c^\dagger_{-\bk \downarrow}
\end{pmatrix},
~ ~ ~ ~ 
{\bf \Psi}^\dagger_\bk =
\begin{pmatrix}
c^\dagger_{\bk\uparrow} & c_{-\bk\downarrow}
\end{pmatrix},
\end{equation}
where $c_{\bk \sigma}$ is the annihilation operator of an electron with the momentum $\bk$ and spin $\sigma$.
Then, the generalized Green's function in the ($2\times 2$) matrix form reads
\begin{equation}
\hat{G}\left(\bk,\tau\right) = - \langle T_\tau {\bf \Psi}_\bk(\tau) {\bf \Psi}_\bk^\dagger \rangle ,
\end{equation}
or using~\eqref{psi_def}, we can write it as
\begin{equation}
\label{EqGreeenExpl}
\hat{G}\left(\bk,\tau\right) =  -
\begin{pmatrix}
\langle T_\tau c_{\bk\uparrow}(\tau)c^\dagger_{\bk\uparrow} \rangle & \langle T_\tau c_{\bk\uparrow}(\tau) c_{-\bk\downarrow} \rangle \\ 
\langle T_\tau c^\dagger_{-\bk\downarrow}(\tau) c^\dagger_{-\bk\downarrow} \rangle & \langle T_\tau c^\dagger_{-\bk\downarrow}(\tau)c_{-\bk\downarrow} \rangle
\end{pmatrix}.
\end{equation}
%
%
The diagonal terms in Eq.~\eqref{EqGreeenExpl} represent the standard Green's functions of electron quasiparticles, 
whereas the off-diagonal terms are the Gor'kov's anomalous Green's functions.
Performing the Fourier transform from the imaginary time domain to the Matsubara frequency representation, we have 
\begin{equation}
\hat{G}(\bk,ip_k) = 
\begin{pmatrix}
\cg(\bk,ip_k) & \cf(\bk,ip_k) \\
\cf^*(\bk,ip_k) & -\cg(-\bk,-ip_k)
\end{pmatrix}.
\end{equation}

To define the bogolons' Green's function, we first introduce the notation $A_\mathbf{p}=u_\mathbf{p}b_{\mathbf{p}}+ v_{\mathbf{p}}b^\dagger_{-\mathbf{p}}$.
Then, the normal and anomalous free Green's functions of the bogolons read $\mathcal{D}(\bp,\tau) = -\langle T_\tau A_\bp(\tau) A^\dagger_\bp \rangle$ and $\mathcal{A}(\bp,\tau) = -\langle T_\tau A_\bp(\tau) A_{-\bp} \rangle$, respectively. 
Switching to the Matsubara frequency domain, we find
\begin{eqnarray}\label{Greenb}
    \mathcal{D}\left(\bp,i\omega_n\right) &=& \frac{u_\bp^2}{i\omega_n-\omega_\bp} - \frac{v_\bp^2}{i\omega_n +\omega_\bp}, \\
    \mathcal{A}\left(\bp,i\omega_n\right) &=& \frac{u_\bp v_\bp}{i\omega_n -\omega_\bp} - \frac{u_\bp v_\bp}{i\omega_n + \omega_\bp}.
\end{eqnarray}
In the long-wavelength limit $p\xi_h \ll 1 $, for the Bogoliubov coefficients~\eqref{eq.4-1} and~\eqref{eq.4-2} we have $u_\bp \approx -v_\bp$, and 
consequently, $\mathcal{D} \left( \bp,i\omega_n\right) \approx - \mathcal{A}\left(\bp,i\omega_n\right)$.
Using the operators $A_\mathbf{p}$ and $A_\mathbf{p}^\dagger$, the bogolon--mediated interaction with the electronic subsystem given in Eqs.~\eqref{Eq1bexpr} and~\eqref{Eq2bexpr} reads
\begin{eqnarray}
&&\mathcal{H}_1 = \frac{\sqrt{n_c}}{L}\sum_{\mathbf{kp}\sigma} g_\bp \left(A_\bp+A_{-\bp}^\dagger\right) c_{\bk+\bp,\sigma}^\dagger c_{\bk,\sigma}, \label{1bexp} \\
&&\mathcal{H}_2 = \frac{1}{L^2} \sum_{\bk\bp\bq\sigma}g_\bp \left( A_{\bq-\bp}^\dagger A_\bq\right)c^\dagger_{\bk+\bp,\sigma}c_{\bk,\sigma}.\label{2bexp}
\end{eqnarray}
Henceforth, we perform a perturbation theory expansion over the weak interacting terms Eqs.~(\ref{1bexp}) and~(\ref{2bexp}) in order to obtain the Dyson equation for the electron Green's function. 
The general solution of the Dyson equation reads
%
\begin{equation}\label{Dyson}
\hat{G}^{-1}(\bk,ip_k) = \hat{G}_0^{-1}(\bk,ip_k) - \hat{\Sigma}(\bk,ip_k),
\end{equation}
%
%
%
%
where
\begin{equation}\label{G0}
\hat{G}_0^{-1}(\bk,ip_k) = i p_k \sigma_0 - \xi_\bk \sigma_3
\end{equation}
is the unperturbed Green's function, $\sigma_{\nu =0,1,2,3}$ are the Pauli matrices, and $\xi_\bk$ is the electron dispersion measured with respect to the chemical potential, $\xi_\bk = \varepsilon_\bk -\mu$.

Following the steps of the standard Eliashberg theory, we separate the electron self--energy into two terms:  the usual Coulomb contribution $\hat{\Sigma}_c$ and the electron--bogolon contribution $\hat{\Sigma}_{eb}$ (in full analogy with the electron-phonon contribution).

The Coulomb contribution is given by
\begin{equation}\label{coulomb}
\hat{\Sigma}_c(\bk,ip_n) = - \frac{1}{\beta} \sum_{\bp,m} \sigma_3 \hat{G}^{od}(\bp,ip_m)\sigma_3 V(\bk-\bp),
\end{equation}
where $V(\bk-\bp)$ are the matrix elements of static screened Coulomb interaction between the electronic states $\bk$ and $\bp$, and $\hat{G}^{od}$ is the off-diagonal component of the Green's function~\cite{AllenBook}.
The self-energies of electron-bogolon ($ \hat{\Sigma}_{1b}$) and electron-bogolon-pair ($\hat{\Sigma}_{2b}$) interaction can be calculated by the Dyson equation up to first order, in accordance with the diagrams presented in Fig.~\ref{fig:diagram},
\begin{widetext}
\begin{eqnarray}
\hat{\Sigma}_{1b}(\bk,ip_k) &=& \sum_{\bp,m} \frac{n_cg_\bp^2}{L^2\beta} \sigma_3\hat{G}(\bk-\bp,ip_k-i\omega_m) \sigma_3 \left[2\mathcal{A}(\bp,i\omega_m) + \mathcal{D}(\bp,i\omega_m) + \mathcal{D}\left(-\bp,-i\omega_m\right)\right], \label{self1b}  \\
\hat{\Sigma}_{2b}\left(\bk,ip_k\right) &=& \sum_{\bp,\bq}^{m,n}  \frac{g_{\bp}^2}{L^4\beta^2} \sigma_3 \hat{G}\left(\bk-\bp,ip_k-i\omega_m -i\omega_n\right) \sigma_3 \\\nonumber
&\times& \left[\mathcal{A} \left(\bq-\bp,i\omega_m\right)\mathcal{A}\left(\bq,i\omega_n\right) +\mathcal{D}\left(\bq-\bp,-i\omega_m\right) \mathcal{D}\left( \bq,i\omega_n \right) \right],  \label{self2b}
\end{eqnarray}
\end{widetext}
where $p_k$ and $\omega_n$ are the Matsubara frequencies of the fermions and bosons, respectively.
Then, in the long-wavelength limit, we see that $\hat{\Sigma}_{1b} \to 0$, because the normal and anomalous Green's functions of the bogolon cancel each other out.
For the 2b processes, we introduce the polarization operator $\mathcal{P}\left(\bp,i\omega_n\right)$, which reads
\begin{widetext}
\begin{eqnarray}
&&\mathcal{P}\left(\bp,i\omega_m\right)=-\frac{2}{\beta L^4} \sum_{\bq,n}\mathcal{A}\left(\bq-\bp,i\omega_n+i\omega_m\right) \mathcal{A} \left(\bq,i\omega_n\right) \label{Pall}\\ \nonumber
&=& -\frac{M^2s^4}{2L^4} \sum_\bq \frac{1}{\omega_{\bq-\bp}\omega_\bq} \left[ \left( \frac{N_{\bq-\bp}- N_\bq}{i\omega_m +\omega_\bq -\omega_{\bq-\bp}} - \frac{N_{\bq-\bp}- N_\bq}{i\omega_m -\omega_\bq + \omega_{\bq-\bp}} \right) + \left( \frac{N_\bq + N_{\bq-\bp}+1}{i\omega_m+ \omega_\bq + \omega_{\bq-\bp}} -\frac{N_\bq + N_{\bq-\bp}+1}{i\omega_m -\omega_\bq-\omega_{\bq-\bp}}\right) \right]. 
\end{eqnarray}
\end{widetext}
In this work, we will restrict ourselves to the case when the contribution of $N_\bq$ terms is negligible. 
As we have shown in earlier works~\cite{Sun21:th,  sun2020boseeinstein}, the $N_\bq$-containing correction only results in quantitative difference (more precisely, it results in an increase of the critical temperature of SC transition), and it is negligible when the size of the condensate is small.
Then, the polarization operator simplifies to
\begin{equation} 
\label{P0}
\mathcal{P}^0\left(\bp,i\omega_m\right) = \frac{-\kappa^2 n_c^2}{L^4} \sum_\bq \frac{1}{\omega_{\bq-\bp} \omega_\bq}\frac{\omega_\bq+\omega_{\bq-\bp}}{\omega_m^2 +\left(\omega_\bq+ \omega_{\bq-\bp}\right)^2},
\end{equation}
and the self-energy can be rewritten in the form,
\begin{eqnarray}
\hat{\Sigma}_{2b}(\bk,ip_k) &=& \frac{-1}{\beta} \sum_{\bp,n} g_{\bk-\bp}^2 \mathcal{P}^0(\bk-\bp,ip_k-ip_n) \nonumber\\ 
&\times&\sigma_3 \hat{G}(\bp,ip_m) \sigma_3. \label{2bself}
\end{eqnarray}
%

We have already demonstrated, that  the self-energy contribution due to 2b process is dominant, $\hat{\Sigma}_{2b} \gg \hat{\Sigma}_{1b}$. 
We want to note, that all other terms like three- and four- (and more) bogolon-mediated processes give smaller contribution since all these terms emerge in the perturbative expansion, where the small parameter is electron-exciton interaction strength $g_\mathbf{p}$.
The terms $\hat{\Sigma}_{1b(2b)}$ are of the same order $g_\mathbf{p}^2$, which is the leading order of the expansion.
%
%

To proceed further, it is a common practice to decompose the matrix self-energy and rewrite it as a linear combination of Pauli matrices with scalar functions as coefficients~\cite{Margine2013},
\begin{eqnarray}\label{Self}
\hat{\Sigma}(\bk,ip_k) &=& ip_k \left[1- Z(\bk,ip_k) \right] \sigma_0 + \chi(\bk,ip_k)\sigma_3 \nonumber \\
&+& \phi(\bk,ip_k) \sigma_1 + \bar{\phi}(\bk,ip_k) \sigma_2,
\end{eqnarray}
where $Z(\bk,ip_k)$ is the mass renormalization function, $\chi(\bk,ip_k)$ is the energy shift, $\phi(\bk,ip_k)$ and $\bar{\phi}(\bk,ip_k)$ are the order parameters.
By the gauge transformation~\cite{AllenBook}, we can set the order parameter $\bar{\phi}$ to zero. 
Then, using~\eqref{Dyson} and~\eqref{G0}, the Green's function reads
\begin{eqnarray}
\hat{G}(\bk,ip_k) &=& -\frac{1}{\Theta(\bk,ip_k)} \Big( ip_k Z(\bk,ip_k)\sigma_0   \nonumber \\
&+& \left[ \xi_\bk + \chi(\bk,ip_k) \right] \sigma_3 + \phi(\bk,ip_k) \sigma_1 \Big), \label{G1} \\
\Theta(\bk,ip_k) &=&  \left[ p_k Z(\bk,ip_k)\right]^2 + \left[\xi_\bk + \chi(\bk,ip_k) \right]^2 \nonumber \\
&+& \left[ \phi(\bk,ip_k) \right]^2. \label{G2}
\end{eqnarray}
Carefully replacing~\eqref{Self} and~\eqref{G1} by the exact forms of self-energy in~\eqref{coulomb} and~\eqref{2bself}, we come up with the Eliashberg equations,
\begin{eqnarray}
Z(\bk,ip_k) &=& 1 + \frac{T}{p_k N_F} \sum_{\bp,n} \frac{p_n Z(\bp,ip_n)}{\Theta(\bp,ip_n)} \lambda(\bk,\bp,k,n),\label{EliaZ} ~ ~ ~ ~ ~\\
\chi(\bk,ip_k) &=& - \frac{T}{N_F} \sum_{\bp,n} \frac{\xi_\bp + \chi(\bp,ip_n)}{ \Theta(\bp,ip_n) } \lambda(\bk,\bp,k,n),\label{EliaChi} \\
\phi(\bk,ip_k) &=& \frac{T}{N_F}  \sum_{\bp,n} \frac{\phi(\bp,ip_n)}{\Theta(\bp,ip_n)} \nonumber \label{EliaPhi} \\
&\times& \left[\lambda(\bk,\bp,k,n) - N_F V(\bk-\bp) \right],
\end{eqnarray}
where $N_F$ is the density of states per spin at the Fermi level and 
\begin{eqnarray} \label{lambda}
\lambda(\bk,\bp,k,n) &=& - N_F g^2_{\bk-\bp} \mathcal{P}^0(\bk-\bp,ip_k-ip_n) \\
&=&\int_0^\infty  \frac{2\omega d\omega}{(p_k- p_n)^2 + \omega^2} \alpha^2F(\bk,\bp,\omega) \nonumber
\end{eqnarray}
is the Eliashberg electron-bogolon spectral function~\cite{AllenBook} with
\begin{equation}
\alpha^2 F(\bk,\bp,\omega) = -\frac{1}{\pi}\mathrm{Im} \left[ \mathcal{P}^0 (\bk-\bp,ip_k-ip_n) \right]. \label{a2F}
\end{equation}
The superconducting gap can now be found as
\begin{equation}
\Delta(\bk,ip_k) = \frac{\phi(\bk,ip_k)}{Z(\bk,ip_k)}.
\end{equation}

For the 2b process, we have
\begin{equation}\label{alphakppmega}
\alpha^2F\left(\mathbf{k},\mathbf{p},\omega \right) =  \frac{ N_F g_{\mathbf{k}-\mathbf{p}}^2 \kappa^2n_c^2}{2L^2}  \frac{\mathtt{H}\left(\omega - s\abs{\mathbf{k-p}}\right)}{2\pi s^2\sqrt{\omega^2 -\left( s\abs{\mathbf{k-p}}\right)^2}},
\end{equation}
where $\mathrm{H}(\omega)$ is the Heaviside step function (see the  details of the calculation in Appendix~\ref{alphaF}).

The equation $\Delta(\bk,ip_k)=0$ gives the normal-state solutions.
The critical temperature $T_c$ can be defined as the highest temperature, for which $\Delta(\bk,ip_k) \neq 0$.
We will use it in what follows.

\section{Results} \label{sec:res}
In this section, we apply the Eliashberg theory to different hybrid systems.
In particular, we consider 2DEG with parabolic dispersion case and the electron gas with the linear dispersion case.

First of all, let us simplify the Eliashberg equations~\eqref{EliaZ}-\eqref{EliaPhi} using the following approximations:
(i) since the superconducting pairing mainly occurs within a narrow energy window around the Fermi surface, we restrict the consideration to the electrons with $\bk_f$~\cite{AllenBook,Margine2013,Schrodi:2020aa}. 
Then, we can put $\chi(\bk_f,ip_k) =0$ and only solve Eqs.~\eqref{EliaZ} and~\eqref{EliaPhi};
(ii) we assume the anisotropy of the Fermi-surface is weak, and thus use the isotropic formulation of the Eliashberg equations.
Then, we can relabel the scalar functions (for brevety):  $Z(\bk=k_f,ip_n) \to Z_n$.


In the parabolic dispersion case (and under the approximations discussed above), Eqs.~\eqref{EliaZ} and~\eqref{EliaPhi} read~\cite{Mahan} (see the details of derivations in the Appendix~\ref{Para_case})
\begin{eqnarray}
\label{EqEliPar1}
Z_n &=& 1+ \frac{\pi T}{p_n} \sum_\nu \frac{p_\nu}{\sqrt{p_\nu^2 + \Delta_\nu^2}} \lambda \left(n-\nu\right) \label{para_Z}, \\
Z_n \Delta_n &=& \pi T \sum_\nu \frac{\Delta_\nu}{\sqrt{p_\nu^2+\Delta_\nu^2}} \left[ \lambda \left(n-\nu\right) -\mu_c^*\right]. \label{para_D}
\end{eqnarray}
Here, we introduce the dimensionless Coulomb interaction $\mu_c^*$. 
By definition, it represents a double-average over the Fermi-surface (FS) of the term $V(\bk-\bp)$ in~\eqref{EliaPhi},
\begin{equation}
\mu_c^* = N_F \langle \langle  V(\bk-\bp) \rangle \rangle_{FS}.
\end{equation}
%
For large class of superconductors~\cite{AllenBook,PhysRev.125.1263,Margine2013}, $\mu_c^*$ is in the range $0.1 \sim 0.2$.
We will use this value instead of calculating the double-average term.

The $\lambda$-function in Eq.~\eqref{EqEliPar1} reads
%
\begin{eqnarray}
&&\lambda \left(n-\nu \equiv m\right) = \frac{\lambda_m}{\sqrt{1+m^2 b_E^2}} \label{ffun} ,~ ~ ~ ~ ~ ~ ~ ~\\
&&\lambda_m = \frac{M^2s N_F}{16\pi k_f } \left( \frac{e_0^2 d}{\epsilon_0 \epsilon}\right)^2 \nonumber \\
&& ~ ~ ~ ~ ~\times \mathcal{F}\left(\arccos \phi_0, \frac{1}{\sqrt{1+m^2 b_E^2}}\right), \label{lambdafun}\\
&& b_E= \frac{\pi T}{sk_f}, ~ ~ ~ ~ \phi_0 = \frac{1}{2k_f L}, \label{const}
\end{eqnarray}
where $m$ is a an integer number, which indicates the difference between two Matsubara frequencies; $\mathcal{F}(\phi,m)$ is the Elliptic integral of the first kind; the density of state per spin is $N_F= m_e/(2\pi)$ with the effective mass of electron $m_e$; and $L$ is the effective size of the condensate introduced as the necessary cut-off. 

In order to calculate $T_c$, we will employ the technique discussed in~\cite{AllenBook,Mahan}.
Since the transition temperature is defined as the point, when the energy gap is infinitesimally small, the value of $T_c$ can be found by setting $\Delta =0$ in all of the denominators of Eqs.~\eqref{para_Z} and~\eqref{para_D}.
This gives
\begin{eqnarray}
&&Z_{2n+1} = 1 + \frac{1}{2n+1} \sum_v \mathrm{sgn}(2v + 1) \lambda(n-v), \label{Z}\\
&&Z_{2n+1} \Delta_{2n+1} = \sum_v \frac{\Delta_{2v+1}}{\abs{2v+1}} \left[ \lambda(n-v) - \mu_c^* \right], \label{ZD}
\end{eqnarray}
where $n$ and $v$ are the Matsubara frequencies' indices, $ip_n = \pi(2n+1)T$, and $\mathrm{sgn}(x)$ is the signum function.
Now, the equations for $Z_{2n+1}$ is independent of the gap function $\Delta_{2n+1}$.
The critical temperature can be found by putting the determinant of the matrix to zero, $\det{M_{nv}}=0$, where %
\begin{equation}
M_{nv} = \delta_{nv} Z_{2n+1} - \frac{\lambda(n-v) - \mu_c^*}{ \abs{2v+1} } .
\end{equation}
Furthermore, we can calculate the $T_c$ numerically. 
%



Let us now consider a 2DEG with a linear dispersion in, e.g., a graphene layer~\cite{Novoselov666,Castro-Neto:2009aa,Novoselov10451,MorozovNature}.
For simplicity, we will disregard the spinor structure of the wave function assuming that we deal with the doped graphene with the Fermi energy sufficiently away from the Dirac point.
Furthermore, we will disregard the contribution from different valleys~\cite{RefEfetov, Kopnin2011}. This approximation is valid since the large non-zero wavevector ($\sim k_f$) strongly depresses the interaction $g_\mathbf{p}$ in Eq.~\eqref{gp}.

Given the assumptions discussed above, we come up with a similar system of equations as the one for the parabolic case: Eqs.~\eqref{para_Z} and~\eqref{para_D}.
However, we find different formula for the coupling constant~\eqref{lambdafun} (see Appendix~\ref{linearApp}),
\begin{equation} \label{lambda_linear}
\lambda_m = \frac{M^2 s}{32 \pi^2 v_f}\left( \frac{e_0^2 d}{\epsilon_0 \epsilon} \right)^2 \mathcal{F} \left( \arccos \phi_0, \frac{1}{\sqrt{1+ m^2 b_E^2}}\right),
\end{equation}
where the definitions of $b$ and $\phi_0$ are same as in Eq.~\eqref{const}.



%
%
%
\begin{figure}[b!]
\centering
\includegraphics[width = 0.47\textwidth]{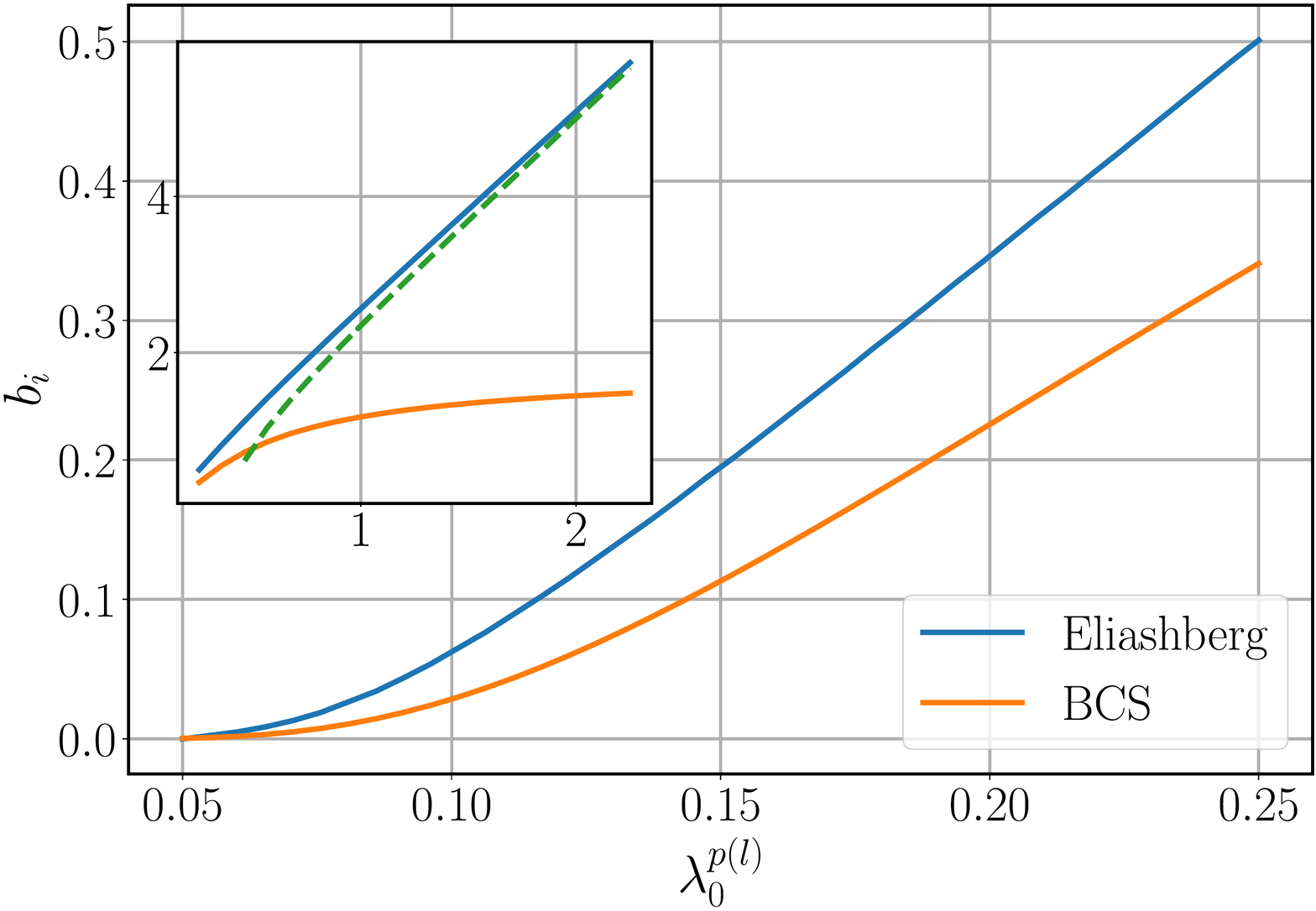}
\caption{Dimensionless critical temperature $b_i$ as a function of the dimensionless coupling constant $\lambda_0^{p(l)}$.
Main plot corresponds to the case of small $\lambda_0^{p(l)}$.
Inset shows the general case of arbitrary $\lambda_0^{p(l)}$. 
Blue line: the results of the calculation by the Eliashberg equations with $b_\textrm{E}=\frac{\pi T}{sk_f}$; 
yellow line: the results of the calculations by the BCS theory with $b_\textrm{BCS} = \frac{\pi T}{2\omega_D}$. Green dashed line in the inset shows the asymptotic estimation with $b_\textrm{E} = \sqrt{\left[\lambda_0^{p(l)} \mathcal{F}(\arccos \phi_0,0.5)\right]^2-1 }$ by Eq.~\eqref{asymptotic}. 
We used the dimensionless parameter $\phi_0 =0.01$. }
\label{fig:dimensionless}
\end{figure}
%
%
%
Let us, first, understand the principle dependence of the critical temperature on the coupling constant.
Comparing Eqs.~\eqref{ffun}, \eqref{lambdafun}, \eqref{const} and~\eqref{lambda_linear}, we see that it is convenient to denote the following dimensionless parameters,
\begin{eqnarray}
\lambda_0^p &=& \frac{M^2s N_F}{16\pi k_f } \left( \frac{e_0^2 d}{\epsilon_0 \epsilon}\right)^2, \label{lambda_0_p} \\
\lambda_0^l &=& \frac{M^2s}{32\pi^2 v_f } \left( \frac{e_0^2 d}{\epsilon_0 \epsilon}\right)^2. \label{lambda_0_l} 
\end{eqnarray}
Then, we can investigate the critical temperature in terms of $b_E=\pi T/sk_f$ and $\lambda_0^{p(l)}$ (Fig.~\ref{fig:dimensionless}).
Let us compare the critical temperatures calculated using the Eliashberg and BCS theories.
\begin{figure}[!t]
\centering
\includegraphics[width = 0.47 \textwidth]{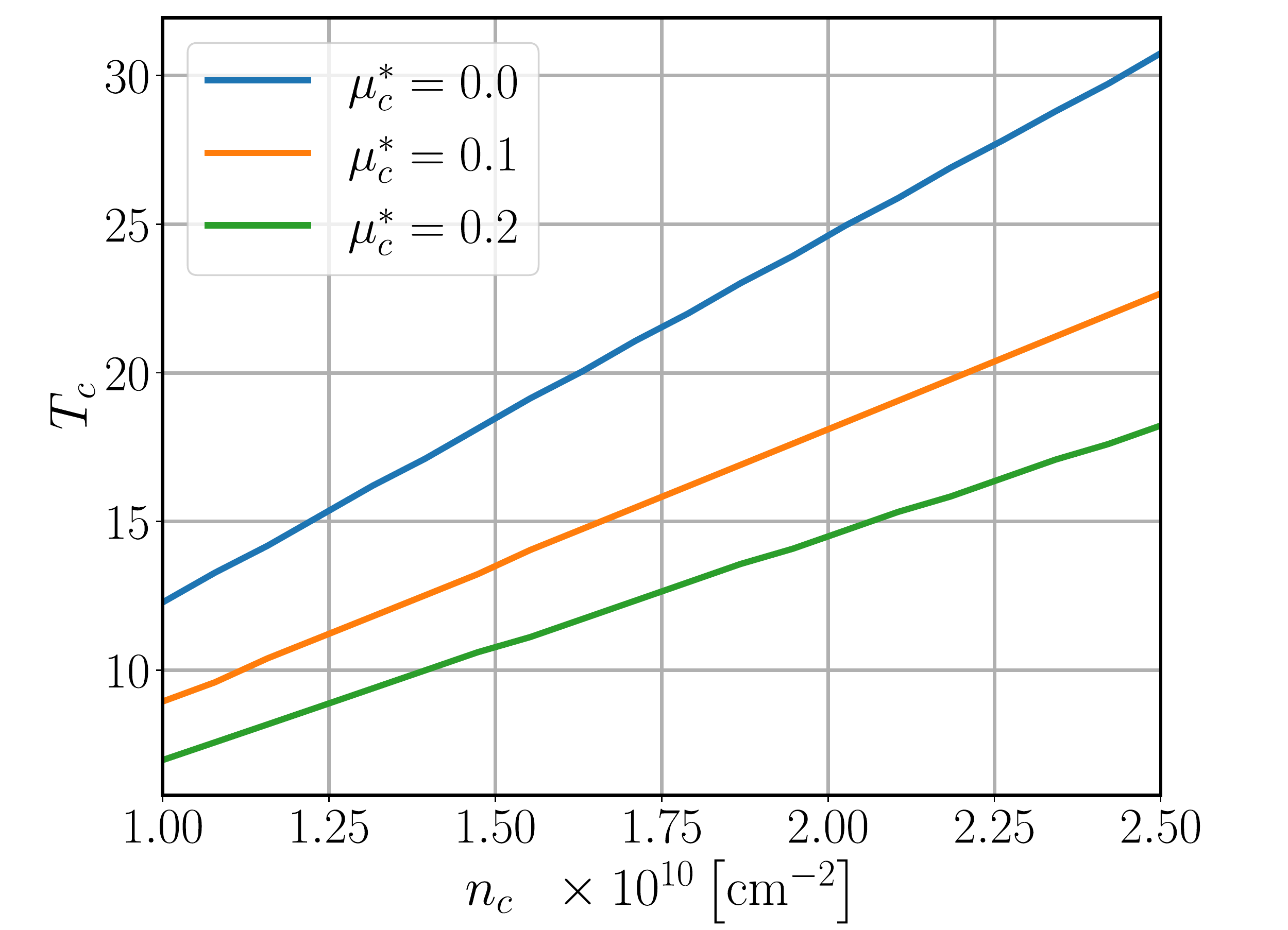}
\caption{The critical temperature of SC transition as a function of condensate density calculated using the Eliashberg theory for different values of dimensionless Coulomb interaction strength $\mu_c^*$. 
We used the parameters typical for MoS$_2$: the electron effective mass $m_e = 0.46m_0$~\cite{RefKormanyos} and exciton effective mass $M=m_0$ with the free electron mass $m_0$; 
the dielectric permittivity $\epsilon = 4.89$; the interlayer separation $d=1.0$~nm, and the electron density $n_e=1.5\times 10^{12}$~cm$^{-2}$.}
\label{fig:TcPara}
\end{figure}
\begin{figure}[t!]
\centering
\includegraphics[width = 0.47\textwidth]{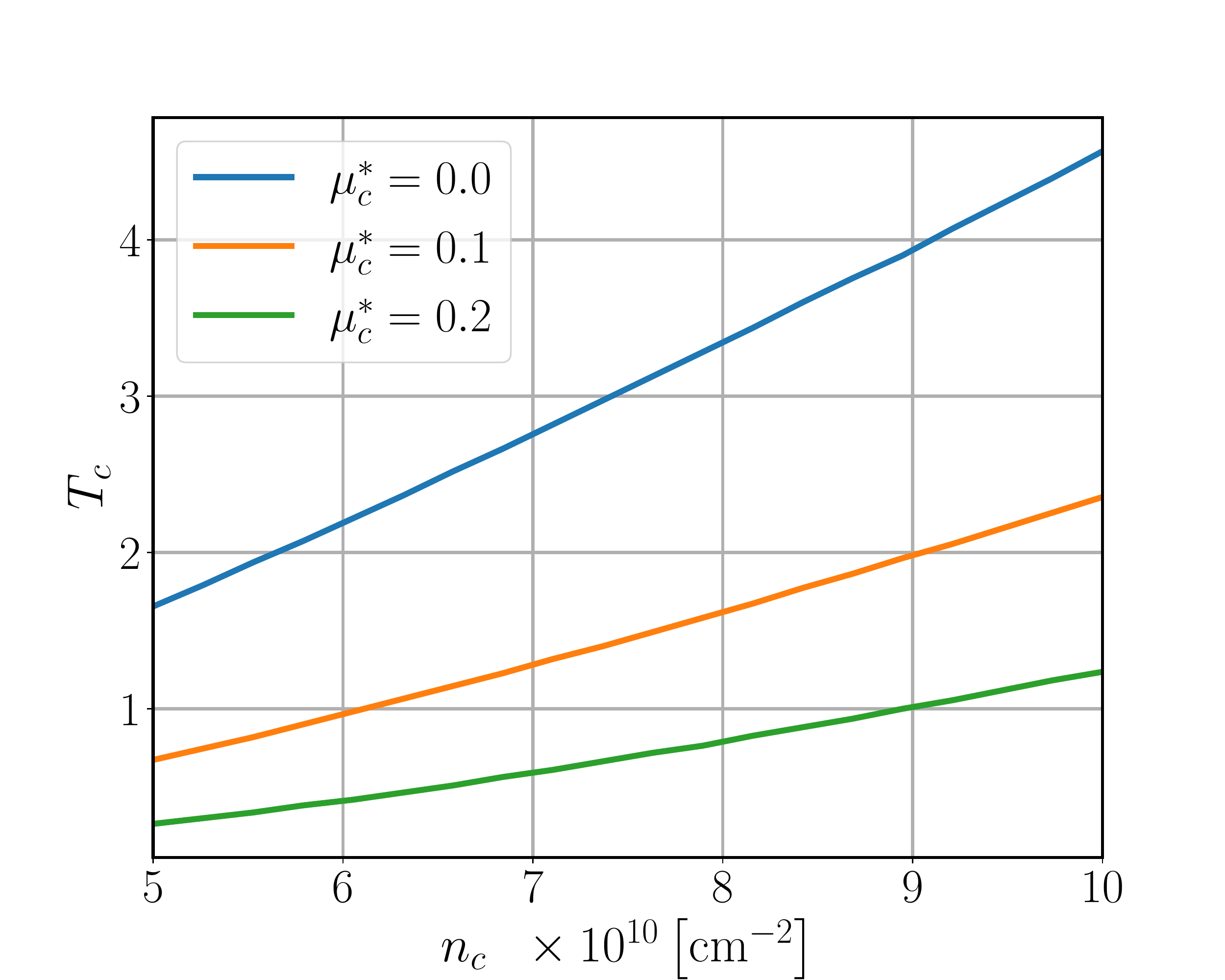}
\caption{Critical temperature as a function of condensate density for the linear dispersion case for different $\mu^*_c$. 
We used $v_f = 10^8$~cm/s. Other parameters are the same as in Fig.~\ref{fig:TcPara}.}
\label{fig:TcLine}
\end{figure}
In BCS, it reads 
\begin{equation}
T_c^\textrm{BCS} = \frac{\gamma}{\pi} 2\omega_D \exp \left( - \frac{1}{\chi}\right),
\end{equation}
where $\omega_D = Ms^2/2$ is the frequency cut-off, and $\gamma = \exp C_0$ with $C_0 \approx 0.577$ the Euler's constant.
The parameter $\chi$ in parabolic~\cite{Sun21:th} and linear~\cite{sun2020boseeinstein} cases reads $\chi_{p(l)} = \lambda_0^{p(l)}\log \left(2 \phi_0^{-1} \right)/\pi$. 
Thus, in BCS we have a different dimensionless temperature, $b_\mathrm{BCS}=\frac{\pi T}{2\omega_D}$ (compare with $b_\mathrm{E}$). 

Figure~\ref{fig:dimensionless} shows that the Eliashberg and BCS theory curves converge well when $\lambda_0^{p(l)}$ is small, as expected; with the increase of $\lambda_0^{p(l)}$, then discrepancies between the two theories become more and more pronounced.
As $\lambda_0^{p(l)} \to \infty$, the asymptotic limit gives a nearly linear dependence of the Eliashberg critical temperature on the coupling coefficient $\lambda_0^{p(l)}$. 
Such dependence is not typical for the conventional superconductors discussed in the early works~\cite{AllenBook, Mahan}, in which the dependence $T_c \sim \sqrt{\lambda}$ in the framework of the Einstein model was found.

The nearly-linear dependence which we find can be also checked analytically if we only consider the leading order of~\eqref{Z} and~\eqref{ZD} (see the details in the Appendix~\ref{asymptotic_limit}).
The asymptotic behaviour reads
\begin{equation} \label{asymptotic}
    b_E = \sqrt{\left[\lambda_0^{p(l)} \mathcal{F}(\arccos \phi_0,b_E)\right]^2-1 }.
\end{equation}
Since the Elliptic integral converges faster as one increases $b_E$, we can treat $\mathcal{F}(\arccos \phi_0,b_E)$ as a constant.   
The inset plot in Fig.~\ref{fig:dimensionless} shows an estimation of the asymptotic behavior of the dimensionless critical temperature with fixed $\mathcal{F}(\arccos \phi_0, b_E=0.5)$.

Figure~\ref{fig:TcPara} and Fig.~\ref{fig:TcLine} show the critical temperature of SC transition as a function of condensation density. 
For the parabolic dispersion, we use  the parameters typical for $\mathrm{MoS}_2$, and for the linear dispersion case we use the parameters typical for graphene. 
Another interesting question is the dependence of electron-bogolon interaction constant on $n_c$ and $n_e$.
In the zeroth order of the electron-bogolon interaction constant, from Eqs.~\eqref{lambda_0_p} and~\eqref{lambda_0_l} we find
\begin{eqnarray}
\lambda_0^{p} &\propto& \sqrt{ \frac{n_c}{n_e}}, \\
\lambda_0^{l} &\propto& \sqrt{n_c}.
\end{eqnarray}
This discrepancy in the parametric dependencies can be understood by recalling the difference between the densities of states in linear and parabolic cases.

Let us address the remaining assumptions, which we used in the calculations.
First, to derive Eqs.~\eqref{lambdafun} and~\eqref{lambda_linear}, in addition to the standard approximations discussed in  Sec.~\ref{sec:res}, we also approximated the exciton-electron interaction by $g_p \approx \frac{e_0^2 d}{2 \epsilon_0 \epsilon}$.
Such a simplification is valid if $k_f d$ and $k_fl$ $\ll 1$.
This assumption imposes a restriction on the maximal allowed value of $n_e$ for considered distances $d$ and $l$.
If the separations $d$ and $l$ are in a nanometre scale, the density of electrons should be $n_e \leq 10^{13}$~cm$^{-2}$.

Second, the polarization operator which we consider in Eq.~\eqref{P0} does not include the $N_\bq$ terms.
These terms are relatively small when the size of the condensate $L_{BEC}$ is small, as it has been shown in the analysis in the framework of the BCS theory~\cite{Sun21:th, sun2020boseeinstein}, since these contributions give an integral truncation from $L_{BEC}^{-1}$ to $k_f/2$.
In other words, we find the lower bound for the SC gap and the $T_c$.
For electrons, we considered the Green's function with account of the finite temperature since we are interested in the critical temperature of the SC transition.
As the earlier works on bogolon~\cite{Sun21:th,sun2020boseeinstein} and phonon~\cite{Enaki_2002}-mediated SC point out, in the BCS theory the non-zero terms in~\eqref{Pall} contribute to a non-monotonous temperature dependence, and they only increase the critical temperature. 
Thus, the simplifications we use are of the technical nature, and they do not qualitatively change our results or conclusions.


\section*{Conclusions}
We developed the Eliashberg theory of the Bose condensate-mediated superconductivity in hybrid two-dimensional Bose-Fermi systems, and showed that bogolon-pair-mediated pairing of electrons represents the dominant mechanism not only in weak but also in the strong-coupling regime, while single-bogolon pairing is suppressed and thus, does not play any significant role.
We started with an analytic expression for the self-energy of electrons in a two-dimensional material due to their interaction with bogolons;
then, we calculated the electron-bogolon coupling constant in the cases of parabolic and linear electron dispersions, and presented the corresponding estimations of the critical temperature of superconducting transition, which turns out relatively high.
It was demonstrated, that the critical temperature of the superconducting transition depends on the dimensionless coupling constant linearly, which is not common for the conventional superconductors.
We expect our theory and the estimations to impact such research areas as low-dimensional superconductors, novel two-dimensional Dirac materials, and the density functional theory for mesoscopic superconductivity.


\begin{acknowledgments}
We thank Prof.~Sergej Flach and Dr.~Kristian Villegas for useful discussions.
We have been supported by the Institute for Basic Science in Korea (Project No.~IBS-R024-D1) and the Ministry of Science and Higher Education of the Russian Federation (Project No.~075-15-2020-797 (13.1902.21.0024)).
\end{acknowledgments}


%
%
%
\begin{widetext}

\appendix
\section{The Eliashberg spectral function} 
\label{alphaF}
Using the definitions Eq.~\eqref{P0} and Eq.~\eqref{a2F}, we find the Eliashberg spectral function~\cite{AllenBook},

\begin{eqnarray} \label{alpha2}
\alpha^2 F(\mathbf{k},\mathbf{p},\omega) &=& N_F g_{\mathbf{k}-\mathbf{p}}^2 \left( -\frac{\kappa^2n_c^2}{2L^2} \right) \int\frac{d\mathbf{q}}{(2\pi)^2}
\frac{ -\delta(\omega-\omega_{\mathbf{q}+\mathbf{k}-\mathbf{p}}-\omega_{\mathbf{q}}) }{ \omega_{\mathbf{q}+\mathbf{k}-\mathbf{p}}\omega_\mathbf{q}}  \nonumber \\
&=& N_F g_{\mathbf{k}-\mathbf{p}}^2 \left( -\frac{\kappa^2n_c^2}{2 L^2} \right) {\cal I}(\omega,\abs{\mathbf{k}-\mathbf{p}}).
\end{eqnarray}
(One can easily verify Eq.~\eqref{alpha2} by performing the integration over $\omega$ in Eq.~\eqref{lambda}.)
Let us take the integral over $\mathbf{q}$.
For that, we denote $\mathbf{p'}=\mathbf{k}-\mathbf{p}$ and then deal with the integral,
\begin{equation}
{\cal I}(\omega,\mathbf{p}')=\int\frac{d\mathbf{q}}{(2\pi)^2} \frac{ (-1) }{ \omega_{\mathbf{q}+\mathbf{p'}}\omega_\mathbf{q}}  \delta(\omega-\omega_{\mathbf{q}+\mathbf{p'}}-\omega_{\mathbf{q}}).
\end{equation}
By denoting $q_1=|\mathbf{q}+\mathbf{p'}|$ (which we will use instead of the angle integration variable below), we find
\begin{equation}
{\cal I}(\omega,\mathbf{p}') = \frac{4}{(2\pi)^2 s^3}\int\limits_0^{\infty }  dq \int\limits_{\abs{q-p'}}^{ q+p'} dq_1\frac{(-1)\delta(\frac{\omega}{s}-q_1-q)}{\sqrt{q_1^2-({ q}-{ p'})^2}\sqrt{({ q}+{ p'})^2-q_1^2}}. 
\end{equation}
Now, we switch the variables of integration using the substitutions $x=s(q+q_1)$ and $y=s(q_1-q)$,  yielding
\begin{eqnarray} \label{Iintegral}
{\cal I}(\omega,\mathbf{p}') &=& \frac{2}{(2\pi)^2 s^2}\int\limits_{sp'}^{\infty }  \frac{dx}{\sqrt{x^2-s^2p'^2}} \int\limits_{-sp'}^{sp'} \frac{dy} {\sqrt{s^2p'^2-y^2}} (-1)\delta(\omega -x)\\ \nonumber
&=& \frac{2}{(2\pi)^2\hbar^3s^2} \frac{(-1)\mathtt{H}(\omega-\hbar sp')}{\sqrt{(\omega/\hbar)^2-s^2p'^2}} \int\limits_{-1}^{1} \frac{dz} {\sqrt{1-z^2}} \\ \nonumber
&=&-\frac{\mathtt{H}\left(\omega -\hbar sp'\right)}{2\pi \hbar^2 s^2\sqrt{\omega^2 -\left(\hbar sp'\right)^2}},
\end{eqnarray}
where $\mathtt{H} \left(x\right)$ is the Heaviside step function.
Substituting Eq.~\eqref{Iintegral} in~\eqref{alpha2}, we find the spectral function given in Eq.~\eqref{alphakppmega}.

\section{The parabolic dispersion case} \label{Para_case}
Using the Eliashberg spectral function~\eqref{alphakppmega}, we find
\begin{eqnarray}
\lambda\left(\mathbf{k},\mathbf{p},n,v\right) &=& \int_0^\infty d\omega \frac{2\omega}{\left( p_n-p_v\right)^2 + \omega^2} N_F g_\mathbf{k-p}^2 \frac{\kappa^2n_c^2}{2L^2} \frac{\mathtt{H}\left(\omega - s\abs{\mathbf{k-p}}\right)}{2\pi  s^2\sqrt{\omega^2 -\left( s \abs{\mathbf{k-p}} \right)^2}} \\
&=&\frac{N_F \kappa^2 n_c^2}{2\pi s^2 L^2}\int_{ s\abs{\mathbf{k-p}}}^\infty d\omega \frac{\omega g_\mathbf{k-p}^2}{\left(p_n - p_v \right)^2+\omega^2} \frac{1}{\sqrt{\omega^2 - \left(  s\abs{\mathbf{k-p}}\right)^2}}. \nonumber
\label{f_t_zero}
\end{eqnarray}
Again, using the notation $\mathbf{q}=\mathbf{k-p}$ and noticing that the integral over $\mathbf{q}$  only depends on the absolute value of $\mathbf{q}$, we find
\begin{eqnarray}
\lambda\left(\mathbf{k},\mathbf{p},n,v\right) &=& \frac{N_F \kappa^2 n_c^2 \abs{g_\mathbf{q}}^2}{4\pi \hbar^2 s^2 L^2} \int_{\left( \hbar s q\right)^2}^\infty \frac{d \omega^2}{\left( p_n -p_v\right)^2 
+\omega^2}\frac{1}{\sqrt{\omega^2 -\left(\hbar s q\right)^2}}\\ \nonumber
&=& \frac{N_F \kappa^2 n_c^2 \abs{g_\mathbf{q}}^2}{4\pi \hbar^2 s^2 L^2} \int_0^\infty \frac{d\Omega}{\left( p_n -p_v\right)^2+\left( \hbar s q\right)^2 +\Omega}\frac{1}{\sqrt{\Omega}}.
\end{eqnarray}
This integral is analytical (for $ab>0$)~\cite{jeffrey2006table}, 
\begin{equation}
\int \frac{dx}{\left( a+ bx\right) \sqrt{x}} = \frac{2}{\sqrt{a b}} \arctan \sqrt{\frac{bx}{a}}.
\end{equation}
Then, denoting $\alpha_{nv} \equiv p_n-p_v$ we find
\begin{equation}
\lambda \left( \mathbf{k},\mathbf{p},n,v \right) = \frac{N_F \kappa^2 n_c^2 g_{\mathbf{k-p}}^2}{4 s^2 L^2 \sqrt{\alpha^2_{n\nu}+\left( s \abs{\mathbf{k-p}}\right)^2}} . 
\end{equation}

Furthermore, under the assumption of isotropic Fermi surface, and considering the pairing to occur on the Fermi surface, from Eq.~\eqref{EliaZ} we find
\begin{eqnarray}
Z_n = 1+ \frac{T}{p_n N_F} \sum_{\mathbf{p},v} \frac{p_v Z_v}{\left[p_v Z_v\right]^2 + \xi_\mathbf{p}^2 + \phi^2_v}  \lambda\left(\mathbf{k},\mathbf{p},n,v\right).
\end{eqnarray}
Taking $k=k_f$, we have
\begin{eqnarray} \label{AppP_Z}
Z_n &=& 1 + \frac{T}{p_n N_F}\sum_v \frac{L^2}{\pi^2}\int dp \frac{p Z_{v} p_v}{p_v^2Z_v^2 + \xi_p^2 + \phi_{v}^2} \int_{k_f-p}^{k_f+p} dq \frac{q}{\sqrt{q^2-\left(k_f-p\right)^2}\sqrt{\left(k_f+p\right)^2 -q^2}}\\ \nonumber
&&\times\frac{N_F \kappa^2 n_c^2 g_{q}^2}{4 s^2 L^2 \sqrt{\alpha_{n\nu}^2+\left( s q\right)^2}},
\end{eqnarray}
where $\mathbf{q}= \mathbf{k}_f-\mathbf{p}$. 
Substituting $p = \sqrt{2m_e\left(\mu+\varepsilon \right)}$, we get
\begin{eqnarray}
Z_n &=& 1 + \frac{T}{p_n}\sum_v \int d\varepsilon \frac{m_e p_v Z_{v}}{p_v^2 Z_{v}^2 + \xi_p^2 +\phi_{v}^2} \frac{1}{\sqrt{q^2 -\left(k_f - \sqrt{k_f^2 +2m_e \varepsilon}\right)^2} \sqrt{\left(k_f + \sqrt{k_f^2+2m_e \varepsilon}\right)^2-q^2}} \nonumber \\
&&\times \frac{L^2}{\pi^2}\int_{k_F-p}^{k_F+p} dq\frac{q \kappa^2 n_c^2 g_q^2}{4s^2L^2\sqrt{\alpha^2_{nv}+ s^2q^2}} \label{Zs1} \\
&\approx& 1 + \frac{T}{p_n} \sum_v \int d\varepsilon \frac{p_v Z_{v}}{p_v^2 Z_{v}^2 + \xi_p^2 +\phi_{v}^2} \frac{1}{q \sqrt{4k_f^2-q^2}} \int_{k_f-p}^{k_f+p} dq \frac{q m_e \kappa^2 n_c^2 g_q^2}{4s^2\pi^2\sqrt{\alpha_{nv}^2 + s^2 q^2}}, \label{Zs3}
\end{eqnarray}
where switching from Eq.~\eqref{Zs1} to~\eqref{Zs3} we made the perturbative expansions using $2m_e\varepsilon \ll k_f^2$.

Furthermore, we can apply the residue theorem to take the integral over  $\varepsilon$.
Also, in the integral over $q$ we change the integral limit to $p=k_f$,
\begin{eqnarray}
\label{EqZinter10}
Z_n &=& 1 + \frac{T}{p_n} \sum_v \frac{\pi p_v} {\sqrt{p_v^2 +\Delta_v^2}}\int_0^{2k_f} \frac{m_e \kappa^2 n_c^2 g_q^2}{4s^2\pi^2 \sqrt{\alpha_{nv}^2+ s^2q^2}} \frac{dq}{\sqrt{4k_f^2-q^2}} \\ \nonumber 
&=& 1 + \frac{T}{p_n}\sum_\nu \frac{p_v}{\sqrt{p_v^2 + \Delta_v^2}} \int_0^{2sk_f} dx \frac{m_e \kappa^2 n_c^2 g_q^2}{4s^2 \pi} \frac{1}{\sqrt{\alpha^2_{nv}+ x^2}}\frac{1}{\sqrt{4s^2k_f^2-x^2}} \\ \nonumber
&\approx& 1 + \frac{T}{p_n}\sum_\nu \frac{p_v}{\sqrt{p_v^2 + \Delta_v^2}} \frac{m_e \kappa^2 n_c^2 g_0^2}{4s^2 \pi} \int_0^{2sk_f} dx  \frac{1}{\sqrt{\alpha^2_{nv}+ x^2}}\frac{1}{\sqrt{4s^2k_f^2-x^2}},
\end{eqnarray}
where the approximation 
\begin{eqnarray}
g_\mathbf{q}^2 = \frac{e_0^4\left(1-\exp\left[-qd \right] \right)^2 \exp\left[-2ql \right]}{4\epsilon^2\epsilon^2_0} 
\approx \frac{e_0^4 d^2}{4\epsilon^2\epsilon^2_0} \equiv g_0^2
\end{eqnarray}
was used.
To proceed with Eq.~\eqref{EqZinter10}, we can use the table integral~\cite{jeffrey2006table}, 
\begin{equation} \label{int1}
\int_u^b \frac{dx}{\sqrt{\left(x^2 + a^2 \right)\left( b^2-x^2\right)}} = \frac{\mathcal{F}\left( \gamma ,r\right)}{\sqrt{a^2 + b^2}}  ~ ~ (b>u\geq0),
\end{equation}
where $\mathcal{F}$ is the elliptic function of the first kind, and
\begin{equation} \label{int1_c}
\gamma = \arccos\frac{u}{b},~ ~ ~ ~ r = \frac{b}{\sqrt{a^2 + b^2}}.
\end{equation}
Finally, 
\begin{equation}
Z_n = 1+ \frac{\pi T}{p_n} \sum_v \frac{p_v}{\sqrt{p_v^2 +\Delta_v^2}} \lambda \left(n-v\right).
\end{equation}

Performing a similar procedure, we obtain Eq.~\eqref{para_D}.


\section{The linear dispersion case} \label{linearApp}
For the linear dispersion case, we start the derivation from Eq.~\eqref{AppP_Z},
\begin{eqnarray}
Z_n &=& 1 + \frac{T}{p_n N_F}\sum_v \frac{L^2}{\pi^2}\int dp \frac{p Z_{v} p_v}{p_v^2Z_v^2 + \xi_p^2 + \phi_{v}^2} \int_{k_f-p}^{k_f+p} dq \frac{q}{\sqrt{q^2-\left(k_f-p\right)^2}\sqrt{\left(k_f+p\right)^2 -q^2}}\\ \nonumber
&&\times\frac{N_F \kappa^2 n_c^2 g_{q}^2}{4 s^2 L^2 \sqrt{\alpha_{n\nu}^2+\left( s q\right)^2}},  
\end{eqnarray}
where $\mathbf{q}= \mathbf{k}_f-\mathbf{p}$ and $p = \pm (\varepsilon+\mu)/v_f$ with $v_f$ the Fermi velocity.
Without the loss of generality, we assume the $n$-doping case of graphene, i.e. $\mu>0$.
Then, the wave vector reads~\cite{kopnin},
\begin{eqnarray}
p = 
\begin{cases}
-\frac{\varepsilon + \mu}{v_f} ~ ~ ~ (\varepsilon<-\mu) \\
\frac{\varepsilon + \mu}{v_f} ~ ~ ~ ~ (-\mu < \varepsilon)
\end{cases},
\end{eqnarray}
and we have 
\begin{eqnarray}
Z_n &=& 1 + \frac{T}{p_n} \sum_v \left[ \int_{-\mu}^\infty  d\varepsilon \left(\frac{\varepsilon+\mu}{v_f^2}\right) \frac{p_v Z_{v}}{p_v^2 Z_{v}^2 + \xi_p^2 +\phi_{v}^2} \frac{1}{\sqrt{q^2 -\left(2k_f + \frac{\varepsilon}{v_f} \right)^2} \sqrt{\left(\frac{\varepsilon }{v_f} \right)^2-q^2}} \right.  \nonumber \\
&& \left. + \int_{-\infty}^{-\mu}d\varepsilon \left( \frac{\varepsilon+\mu}{v_f^2} \right) \frac{ p_v Z_{v}}{p_v^2 Z_{v}^2 + \xi_p^2 +\phi_{v}^2} \frac{1}{\sqrt{q^2 -\left( \frac{\varepsilon }{v_f} \right)^2} \sqrt{\left(2k_f + \frac{\varepsilon }{v_f} \right)^2-q^2}} \right]  \nonumber \\
&&\times \frac{L^2}{\pi^2}\int_{k_F-p}^{k_F+p} dq\frac{q \kappa^2 n_c^2 g_q^2}{4s^2L^2\sqrt{\alpha^2_{nv}+ s^2q^2}}.
\end{eqnarray}
Assuming $\frac{\varepsilon}{v_f} \ll k_f$ and $\frac{\varepsilon}{v_f} \ll q$, we can write
\begin{eqnarray}
Z_n &\approx& 1 + \frac{T}{p_n} \sum_v \left[ \int_{-\infty}^\infty  d\varepsilon \left(\frac{\varepsilon}{v_f^2}\right) \frac{p_v Z_{v}}{p_v^2 Z_{v}^2 + \xi_p^2 +\phi_{v}^2} \frac{1}{\sqrt{q^2 \left( 4k_f^2 - q^2 \right) } } \right.  \nonumber \\
&& \left. + \int_{-\infty}^{\infty} d\varepsilon \left( \frac{\mu}{v_f^2} \right) \frac{ p_v Z_{v}}{p_v^2 Z_{v}^2 + \xi_p^2 +\phi_{v}^2} \frac{1}{\sqrt{q^2 \left(4k_f^2-q^2 \right) }} \right]  \nonumber \\
&&\times \frac{L^2}{\pi^2}\int_{k_f-p}^{k_f+p} dq\frac{q \kappa^2 n_c^2 g_q^2}{4s^2L^2\sqrt{\alpha^2_{nv}+ s^2q^2}}.
\end{eqnarray}
%
The integrand in the first line represents an odd function of $\varepsilon$, and thus it gives a vanishing contribution.
For the rest terms in the equation above, we apply the residue theorem to take the integral over $\varepsilon$.
Also, in the integral over $q$, we change the limit of integration to $p=k_f$,
\begin{eqnarray}
Z_n &=& 1 + \frac{T}{p_n} \sum_v \frac{\pi p_v} {\sqrt{p_v^2 +\Delta_v^2}}\int_0^{2k_f} \frac{\mu \kappa^2 n_c^2 g_q^2}{4s^2\pi^2 v_f^2 \sqrt{\alpha_{nv}^2+ s^2q^2}} \frac{dq}{\sqrt{4k_f^2-q^2}} \\ \nonumber 
&=& 1 + \frac{T}{p_n}\sum_\nu \frac{p_v}{\sqrt{p_v^2 + \Delta_v^2}} \int_0^{2sk_f} dx \frac{ \mu \kappa^2 n_c^2 g_q^2}{4s^2 v_f^2 \pi} \frac{1}{\sqrt{\alpha^2_{nv}+ x^2}}\frac{1}{\sqrt{4s^2k_f^2-x^2}} \\ \nonumber
&\approx& 1 + \frac{T}{p_n}\sum_\nu \frac{p_v}{\sqrt{p_v^2 + \Delta_v^2}} \frac{ \mu \kappa^2 n_c^2 g_0^2}{4s^2 v_f^2\pi} \int_0^{2sk_f} dx  \frac{1}{\sqrt{\alpha^2_{nv}+ x^2}}\frac{1}{\sqrt{4s^2k_f^2-x^2}}.
\end{eqnarray}
Using the relation given in Eqs.~\eqref{int1} and~\eqref{int1_c}, we find
\begin{eqnarray}
Z_n = 1 + \frac{\pi T}{p_n} \sum_v \frac{p_v}{\sqrt{p_v^2 + \Delta_v^2}} \frac{k_f \kappa^2 n_c^2 g_0^2}{4 s^2 \pi^2 v_f} \frac{1}{ \sqrt{\alpha_{nv}^2 + 4s^2 k_f^2} } \mathcal{F}\left( \arccos \frac{1}{2k_f L}, \frac{2sk_f}{ \sqrt{\alpha_{nv}^2 + 4s^2k_f^2}}\right).
\end{eqnarray}
Introducing the quantities
\begin{eqnarray}
&&\lambda\left( n-v \equiv m \right) = \frac{\lambda_m}{\sqrt{1+ m^2 b_E^2}}, \\
&&\lambda_m = \frac{M^2 s}{32 \pi^2 v_f}\left( \frac{e_0^2 d}{\epsilon_0 \epsilon} \right)^2 \mathcal{F} \left( \arccos \phi_0, \frac{1}{\sqrt{1+ m^2 b_E^2}}\right), \\
&& b_E = \frac{\pi T}{sk_f}, ~ ~ ~ ~ ~ \phi_0 = \frac{1}{2Lk_f},
\end{eqnarray}
we come up with a relatively simple final expression for the mass renormalization function, 
\begin{equation}
Z_n = 1 + \frac{\pi T}{p_n} \sum_v \frac{p_v}{\sqrt{p_v + \Delta_v}} \lambda \left( n-v \right).
\end{equation}


%
%
%
\begin{figure}[b!]
    \centering
    \includegraphics[width=0.45\textwidth]{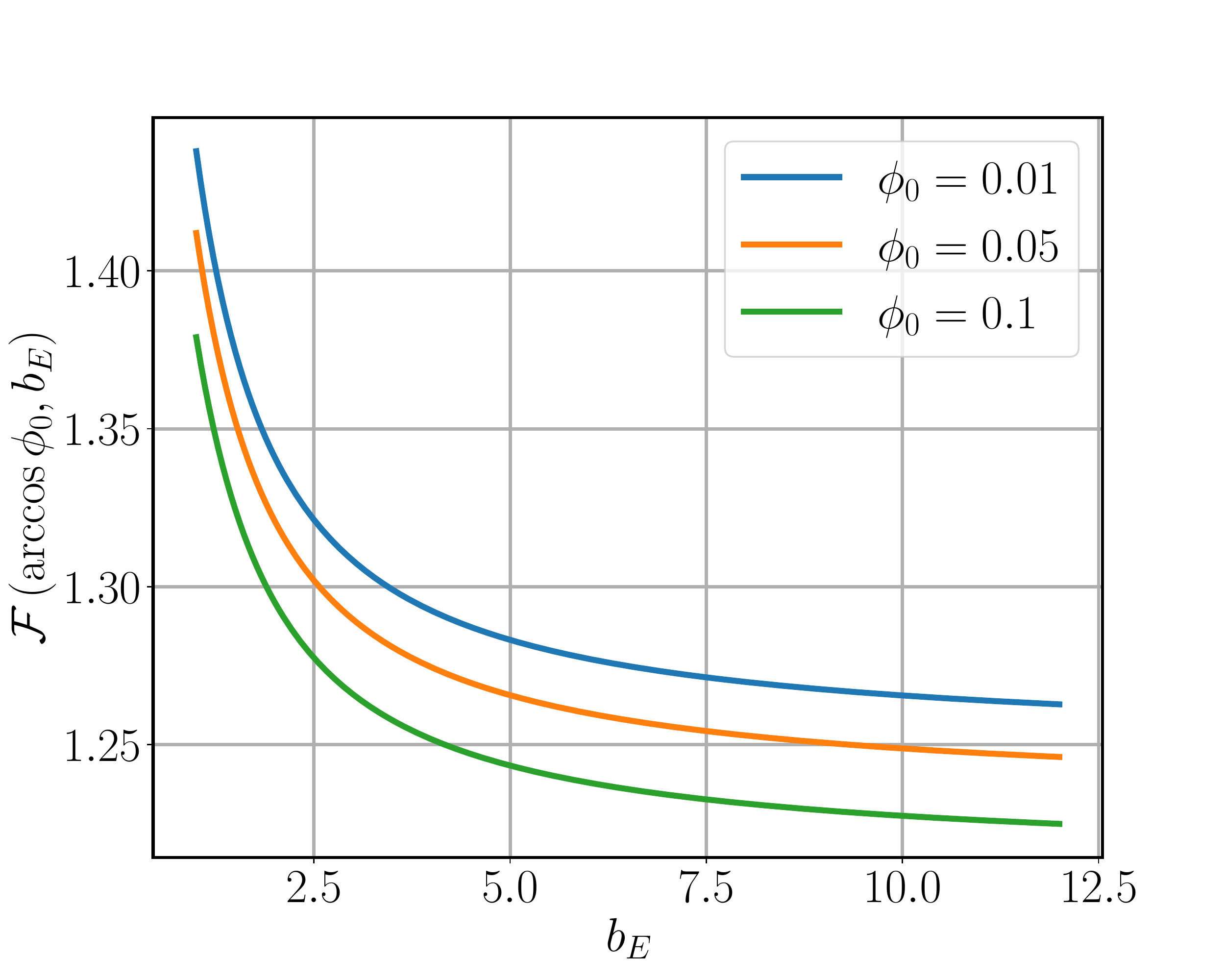}
    \caption{Elliptic integral $\mathcal{F}(\arccos \phi_0, b_E)$ as a function of dimensionless parameter $b_E$.}
    \label{fig:dimensionless_F_B}
\end{figure}

\section{The Asymptotic limit} 
\label{asymptotic_limit}
Let us separately address the asymptotic limit of large $\lambda_0^{p(l)}$.
Following the arguments discussed in~\cite{Mahan}, we can assume $b_E$ to be large in this case.
Then, the leading order contribution of~\eqref{ffun} or~\eqref{lambda_linear} is
\begin{equation}
    \lambda\left(0\right)= \lambda_0^{p(l)} \mathcal{F}\left(\arccos\phi_0 \equiv \Tilde{\phi_0},1\right),
\end{equation}
with the zero Matsubara frequency index, and we have neglected the terms $\lambda(n)$ with $\abs{n}>1$.
Then, Eqs.~\eqref{Z} and~\eqref{ZD} are truncated.
For Eq.~\eqref{Z}, we find $Z_{\pm 1} = 1+\lambda(0)$, and for Eq.~\eqref{ZD} we have $Z_{\pm 1} \Delta_{\pm 1} = \Delta_{\pm 1} \lambda(0) + \Delta_{\mp 1} \lambda(\pm 1)$.
This results in a self-consistent equation,
\begin{equation}
    \left[ \frac{\chi}{\sqrt{1+ b_E^2}} \mathcal{F}\left(\Tilde{\phi_0}, \frac{1}{\sqrt{1+b_E^2}} \right)\right]^2 = 1.
\end{equation}
Furthermore, we can assume $\mathcal{F}(\Tilde{\phi_0}, \frac{1}{\sqrt{1+b_E^2}})$ to be constant for simplicity since it converges to a constant when the temperature increases, as it is shown in Fig.~\ref{fig:dimensionless_F_B}. 
Then, we find the following estimation for $T_c$,
\begin{equation}
    b_E = \sqrt{\chi^2 \mathcal{F}_c^2-1 },
\end{equation}
where the symbol $\mathcal{F}_c$ means we have fixed $b_E$ as a constant inside $\mathcal{F}\left(\Tilde{\phi_0}, \frac{1}{\sqrt{1+b_E^2}}\right)$.
\end{widetext}

\bibliography{library}

\end{document}